# RAY INTERPRETATION OF THE TRANSITION RADIATION PROCESS IN A WAVEGUIDE

Mikayel Ivanyan, Lusine Aslyan
CANDLE, Acharyan str 31, 0040 Yerevan, Armenia

*Abstract*
The paper demonstrates the geometric optic properties of the transition radiation (TR) of an ultrarelativistic point-like charged particle crossing the transverse wall of a circular semi-infinite waveguide with ideally conducting walls. It is shown that the rays forming the TR field emanate from the point of entry and obey the laws of geometric optics. The TR field at the observation point is formed from geometric-optical rays multi re-reflected from the waveguide wall.

As is known, TR in unlimited space, formed by a charged particle crossing the boundary of two media with an oblique incidence, obeys the geometric-optic laws of refraction and reflection of light [1, 2]. Is there a connection between the propagation of radiation and the laws of geometrical optics in the case of transition radiation in a waveguide? In numerous studies on TR in waveguides [3, 4] (we quote the most characteristic of them) this question is not covered, possibly because of the complexity of the formulae describing the TR field in the form of waveguide eigenfunction expansions. In this regard, we can only note the work [5].

In this paper, we use a numerical example to show that the TR field in a waveguide strictly follows the geometric-optical reflection laws. We demonstrate the geometric properties of the transition radiation (TR) of an ultrarelativistic point-like charged particle crossing the transverse wall of a circular semi-infinite waveguide. The particle trajectory is parallel to the waveguide axis (Fig.1). Both the cylindrical wall of the waveguide and the transverse boundary are ideal conductors.

In this case, the solution for TR is obtained in a closed form in the inner region of a semi-infinite waveguide from.

The frequency distribution of the longitudinal electric component of the TR field of ultra-relativistic particle inside the waveguide has the form of ideal waveguide mode series:

$$E_z^{Tr} = \sum_{m=-\infty}^{\infty} \sum_{n=1}^{\infty} jq \frac{Z_0}{\pi a} \frac{J_m(\chi_{mn}\tilde{r}_0)}{\sqrt{\tilde{k}^2-\chi_{mn}^2}} \frac{J_m(\chi_{mn}\tilde{r})}{J_{m+1}^2(\chi_{mn})} exp(jm\varphi) exp\left\{j\left(\tilde{k}\tilde{t} - \sqrt{\tilde{k}_c^2 - \chi_{mn}^2}\tilde{z}\right)\right\} \qquad (1)$$

Here $\tilde{k} = \omega a/c$, $\tilde{z} = z/a$, $\tilde{r} = r/a$, $\tilde{t} = ct/a$ dimensionless wavenumber ($a$ waweguide radii, $\omega$ frequency), normed longitudinal and radial coordinates (in cylindrical coordinate frame $r, \varphi, z$, Fig.1) and dimensionless time, $J_m$ is a Bessel Function of first kind and $m^{th}$ order and $\chi_{mn}$ $n^{th}$ root of $m^{th}$ order Bessel function ($J_m(\chi_{mn}) = 0$); $\{\tilde{r}_0 < a, 0, 0\}$ particle entry point, $\{\tilde{r}, \varphi, \tilde{z}\}$ observation point.

In the time domain:

$$E_z^{Tr}(r, \varphi, z, t) = \int_{-\infty}^{\infty} E_z^{Tr} d\omega = jq \frac{Z_0 c}{\pi a^2} \sum_{m=0}^{\infty} \delta_m \sum_{n=1}^{\infty} J_m(\chi_{mn}\tilde{r}_0) \frac{J_m(\chi_{mn}\tilde{r})}{J_{m+1}^2(\chi_{mn})} Cos\{m\varphi\} G(\tilde{z}, \tilde{t}) \qquad (2)$$

$$G(\tilde{z}, \tilde{t}) = \begin{cases} 2K_0\left[\chi_{mn}\sqrt{\tilde{z}^2 - \tilde{t}^2}\right], & 0 < \tilde{t} < \tilde{z} \\ j\pi H_0^{(1)}\left[\chi_{mn}\sqrt{\tilde{t}^2 - \tilde{z}^2}\right], & \tilde{z} < \tilde{t} < \infty \end{cases} \qquad (3)$$

Here $\delta_0 = 1/2, \delta_{m>0} = 1, H_0^{(1)}, K_0$ zero order Hankel and modified Bessel function of the second kind, respectively. In constructing the decomposition (2, 3), we used the tables of integral transformations [6]. The remaining components of the TR fields can be expressed in terms of derivatives of the longitudinal electrical components (for both frequency (1) and space-time (2, 3) representations):

$$E_\phi^{Tr} = \frac{a^2}{\chi_{mn}^2 r} \frac{\partial^2}{\partial \varphi \partial z} E_z^{Tr}, \quad E_r^{Tr} = \frac{a}{\chi_{mn}} \frac{\partial^2}{\partial r \partial z} E_z^{Tr}, \quad H_\varphi^{Tr} = j \frac{a}{\chi_{mn}} \frac{\partial^2}{\partial t \partial r} E_z^{Tr}, \quad H_r^{Tr} = -j \frac{a}{\chi_{mn}^2 r} \frac{\partial^2}{\partial t \partial \varphi} E_z^{Tr}, \quad H_z^{Tr} = 0 \quad (4)$$

Further actions are as follows: at a fixed observation point inside the waveguide, at a certain distance from the transverse wall, the time dependence of the value of the longitudinal electric component is calculated. The discrete time moments at which the field in observation point is not equal to zero (in these temporary moments, the series (2) diverges) are identified with the optical lengths of the rays connecting the entry point with the observation point. The ray's trajectory are conditioned by their reflections from the walls of the waveguide according to the laws of geometrical optics.

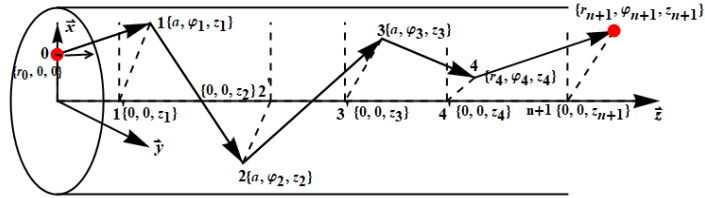

Fig.1. Geometry of the problem. Basic notation

When $\varphi \neq 0$ and $r_0 > 0$, the rays emanating from the point $\{r_0, 0, 0\}$ and, with undergoing a series of geometrical-optical reflections, converging at the point $\{r_{n+1}, \varphi_{n+1}, z_{n+1}\}$, have a complex trajectory. We have developed a method for determining the longitudinal and polar coordinates of the points of reflection from the surface of the waveguide and the selection of various options for the trajectories that meet the stated requirements. We briefly outline the main provisions. Figure 1 schematically shows the course of the rays. A prerequisite for meeting the condition of conformity of the path of the rays to geometrical optics is to find in the same plane the incident and reflected rays and the normal, dropped from the reflection point to the axis of the waveguide, those, four points must be in the same plane (Fig.1): $i-1, i, i'$ and $i+1$, $i = 1,2,\ldots n$, where $n$ is the number of reflections. This requirement reduces to a system of equations in Cartesian frame [7]:

$$Det \begin{cases} \begin{pmatrix} x_{i-1} & y_{i-1} & z_{i-1} & 1 \\ x_i & y_i & z_i & 1 \\ x_{i'} & y_{i'} & z_{i'} & 1 \\ x_{i+1} & y_{i+1} & z_{i+1} & 1 \end{pmatrix} \end{cases} = 0, \quad i = 1,2,\ldots n \quad (5)$$

Here $\{x_0, y_0, z_0\}$ entry point, $\{x_{n+1}, y_{n+1}, z_{n+1}\}$ observation point. The system 5) defines the relationship between the longitudinal $z_i$ and angular $\varphi_i$ coordinates of the reflection points:

$$z_j = z_{n+1} \frac{r_0 \sin(\varphi_1) - \delta_j a \sum_{i=1}^{j-1} \sin(\varphi_i - \varphi_{i+1})}{r_0 \sin(\varphi_1) - r_{n+1} \sin(\varphi_n - \tilde{\varphi}_{n+1}) - \delta_n a \sum_{i=1}^{n-1} \sin(\varphi_i - \varphi_{i+1})}, \quad \delta_1 = 0, \delta_{j>1} = 1, j \leq n \quad (6)$$

Another requirement that follows from the laws of geometric optics is the equality of the angles of incidence and the reflection of rays:

$$\angle\{i-1, i, i'\} = \angle\{i', i, i+1\}, i = 1, 2, \ldots n \tag{7}$$

Substituting relations (6) into the solutions obtained from the system (7), we obtain the equations for the angle $\varphi_1$:

For $n = 1$: $A_0 + \sum_{i=1}^{3}\{A_i \cos(i\varphi_1) + B_i \sin(i\varphi_1)\} = 0$ (8)

with $A_0 = a^2(r_0^2 - r_{n+1}^2)$,

$A_1 = -a r_0 r_{n+1}(r_0 \cos\varphi_{n+1} - r_{n+1}(2 - \cos 2\varphi_{n+1}))$, $B_1 = -a r_0 r_{n+1}(3 r_0 + 2 r_{n+1} \cos\varphi_{n+1}) \sin\varphi_{n+1}$

$A_2 = -r_0^2(a^2 + r_{n+1}^2) + r_{n+1}^2(a^2 + r_0^2)\cos 2\varphi_{n+1}$, $B_2 = r_{n+1}^2(a^2 + r_0^2) \sin 2\varphi_{n+1}$,

$A_3 = a r_0 r_{n+1}(r_0 \cos\varphi_{n+1} - r_{n+1} \cos 2\varphi_{n+1})$, $B_3 = a r_0 r_{n+1}(r_0 - 2 r_{n+1} \cos\varphi_{n+1}) \sin\varphi_{n+1}$;

For $n > 1$: $a^2 + (a^2 + r_n^2)\cos\beta - 4 a r_n \cos^2\frac{\beta}{2} \cos\alpha + r_n^2 \cos 2\alpha = 0$ (9)

$\beta = (-1)^j \text{Arccos}\left(-\frac{a^2 - 2 a r_0 + r_0^2 \cos 2\varphi_1}{a^2 + r_0^2 - 2 a r_0 \cos\varphi_1}\right), j = 0, 1; \alpha = (n-1)\beta - \varphi_1 + \tilde{\varphi}_{n+1}; \varphi_2 = \varphi_1 - \beta.$

For $n > 2$: $\varphi_{i+2} = -\varphi_i + 2\varphi_{i+1}, i = 1, 2, \ldots n-2$.

From equations (8, 9) it follows that the angular coordinates of the reflection points $\varphi_i$ and ratios $\frac{z_j}{z_{n+1}}$ (6) do not depend on the longitudinal coordinate of the observation point $z_{n+1}$. Substituting the polar angle values $\varphi_i$ into (6) determines the longitudinal coordinates $z_j$ of the reflection points on the inner surface of the cylinder. Equations (8, 9) are solved numerically. Now the optical paths of the rays are written as follows:

For $n = 0$ (direct hit of the beam):

$$L_0 = l_0, \text{ for } n > 1: L_n = l_0 + l_n + \delta_n \sum_{i=1}^{n-1} l_i;$$
$$l_i = \sqrt{(x_i - x_{i+1})^2 + (y_i - y_{i+1})^2 + (z_i - z_{i+1})^2} \tag{10}$$

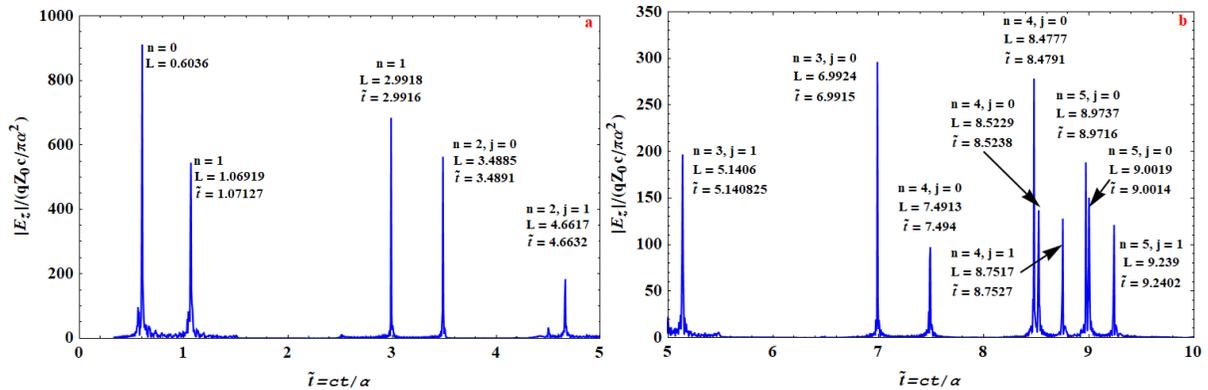

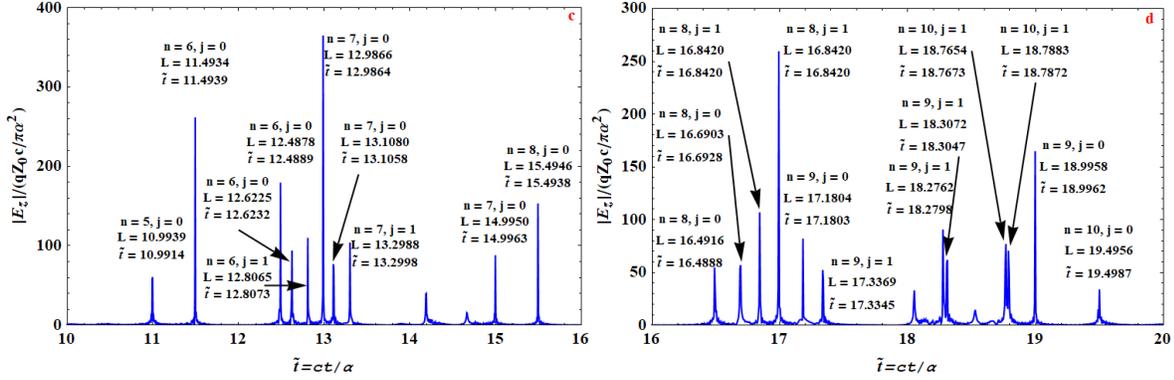

Fig.2. Time dependence of the longitudinal electric component of TR field at a fixed point of observation. Each peak corresponds to the number of reflections (n), the optical path length (L) and the time of the field at the observation point ($\tilde{t}$). Index $j$ determines the sign of the function $\beta$ (9).

In Figure 2, the calculations, made by the formulae (8-10), are compared with the extreme values of the longitudinal electric field component of TR, calculated by the formulae (2, 3). The time dependence of TR field at a fixed point of observation $\{\tilde{r} = 0.25, \varphi = \pi/6, \tilde{z} = 0.25\}$ with the point of particle entry $\{\tilde{r}_0 = 0.75, \varphi = 0, \tilde{z} = 0\}$ was calculated. The temporal distribution of the field is discrete: it is not equal to zero at only at certain points in time. Peaks cannot be attributed to certain waveguide modes: they are formed as a result of the imposition of an infinite number of them. Their occurrence can be explained, if we assume that the TR is an instantaneous pulse arising at the time moment of entry of the particle, or, in other words, the point of entry of a particle is an instantaneous source of radiation, emitting rays inside the waveguide in all directions, but only rays whose optical path corresponds to the laws of geometric optics reach the observation point. The location of the peaks in Figure 2 indicate on the geometric optical nature of their occurrence: as the inscriptions in the Figures indicate, each of the peaks corresponds to a ray emanating from the point of entry of the particle and reaching the observation point after the corresponding number ($n$) of reflections. The optical path ($L$) of the corresponding beam is equal to the time moment ($\tilde{t}$) of formation of the peak (in accordance with the principle of causality). In Fig. 2, the correspondence between the moments of peak formation and the optical paths of rays experiencing up to ten reflections is traced. Thus, our statement about the formation of a TR field in accordance with the laws of geometric optics can be considered as proven.

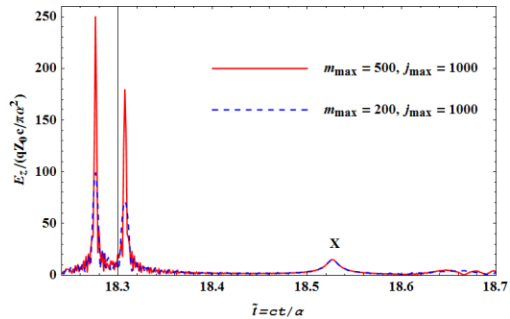

Fig.3. Effect of increasing the number of waveguide modes taken into account. Related to Fig.2, d

In the calculations shown in Fig. 2, a finite number of harmonics were used: $m = 0,1,2,..200; j = 1,2,...1000$. This explains the presence in Figure 2 of several unidentified peaks. When using more harmonics (Fig.3), the peaks due to the geometrical optics increase, and the height of the unidentified peaks remains unchanged (on Fig.3 is marked with a cross).

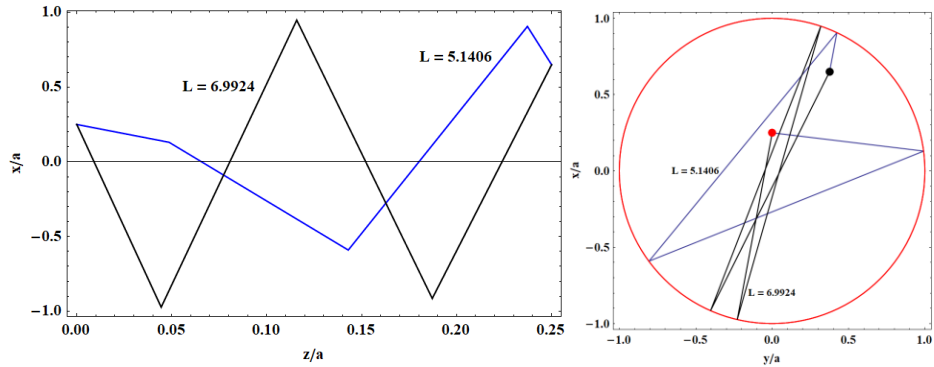

Fig.4. Case of triple reflection. The ray's trajectories projection on the plane x, z (left) and x, y (right); enter point (red), observation point (black). Related to Fig. 2, b

Figure 4 shows an example of the trajectories of rays that have experienced triple reflection from the waveguide wall. In this case, the rays reach the observation point in two different ways: $j = 0$ and $j = 1$. As the number of reflections increases, the number of variations may also increase. So for n=4 there are 4 different trajectories (Fig. 2, b). Over time, the number of reflections increases, but the geometric-optical character of the processes produced by TR, persists.

**Conclusion**

The regularities demonstrated for an ideal waveguide can be extended to both resistive waveguides and waveguides partially filled with the medium. The applied technique can be extended to closed resonators.

**Acknowledgment**

This work was supported by the RA MES State Committee of Science, in the frames of the research project № 17A-1C004